\documentclass[12pt,preprint]{aastex}
\usepackage{amsmath}
\pdfoutput=1

\newcommand{\gapprox}   {\lower.4ex\hbox{$\;\buildrel >\over{\scriptstyle\sim}\;$}}
\newcommand{\lapprox}   {\lower.4ex\hbox{$\;\buildrel <\over{\scriptstyle\sim}\;$}}

\newcommand{\begeq}     {\begin{equation}}
\newcommand{\fineq}     {\end{equation}}

\newcommand{\fouruooneonefive}{\mbox{4U\,0115$+$63}}

\newcommand{\lmcxfour}{\mbox{LMC\,X-4}}
\newcommand{\herxone}{\mbox{Her\,X-1}}
\newcommand{\cenxthree}{\mbox{Cen\,X-3}}

\newcommand{\xper}{\mbox{X\,\,Per}}

\begin{document}

\title{The NuSTAR X-ray Spectrum of Hercules X-1: A 
Radiation-Dominated Radiative Shock}

\author{Michael T. Wolff\altaffilmark{1},
Peter A. Becker\altaffilmark{2},
Amy M. Gottlieb\altaffilmark{3,4},
Felix F\"urst\altaffilmark{5},
Paul B. Hemphill\altaffilmark{6},
Diana M. Marcu-Cheatham\altaffilmark{3,4},
Katja Pottschmidt\altaffilmark{3,4},
Fritz-Walter Schwarm\altaffilmark{7},
J\"orn Wilms\altaffilmark{7}, and
Kent~S.~Wood\altaffilmark{1,8}}
\altaffiltext{1}{Space Science Division, Naval Research Laboratory, Washington, DC 20375-5352, USA}
\altaffiltext{2}{Department of Physics \& Astronomy, George Mason University, Fairfax, VA 22030-4444, USA}
\altaffiltext{3}{Department of Physics \& Center for Space Science and Technology, University of Maryland Baltimore County, Baltimore, MD 21250, USA}
\altaffiltext{4}{CRESST \& NASA Goddard Space Flight Center, Code 661, Greenbelt, MD  20771, USA}
\altaffiltext{5}{Cahill Center for Astronomy and Astrophysics, California Institute of Technology, Pasadena, CA 91125, USA}
\altaffiltext{6}{Center for Astrophysics and Space Sciences, University of California, San Diego, 9500 Gilman Dr., La Jolla, CA 92093-0424, USA}
\altaffiltext{7}{Dr. Karl-Remeis-Sternwarte and ECAP, Sternwartstr, 7, 96049 Bamberg, Germany}
\altaffiltext{8}{Praxis, Inc., 5845 Richmond Hwy., Suite 700, Alexandria, VA 22303  USA}

\begin{abstract}

We report new spectral modeling of the accreting X-ray pulsar 
Hercules X-1.
Our radiation-dominated radiative shock model is an 
implementation of the analytic work of \citeauthor{bw07} on 
Comptonized accretion flows onto 
magnetic neutron stars.
We obtain a good fit to the spin-phase averaged 4 to 78 keV 
X-ray spectrum observed by the Nuclear Spectroscopic Telescope 
Array during a main-on phase of the \herxone\ 35-day 
accretion disk precession period.
This model allows us to estimate the accretion rate, the 
Comptonizing temperature of the radiating plasma, 
the radius of the magnetic polar cap, and the average 
scattering opacity parameters in the accretion column.
This is in contrast to previous phenomenological models that characterized the 
shape of the X-ray spectrum but could not determine the physical 
parameters of the accretion flow.
We describe the spectral fitting details and discuss 
the interpretation of the accretion flow physical parameters.

\end{abstract}

\keywords{radiation: dynamics, X-rays: binaries, stars: individual (\herxone)}

\section{Introduction}
\label{section:introduction}

In the standard model for accreting X-ray pulsars, the matter 
flows from a companion star either via a stellar wind or Roche 
lobe overflow into an accretion disk around a spinning highly 
magnetic neutron star (NS). 
As the orbiting matter approaches the NS, the increased magnetic 
field strength constrains the material to flow along the 
magnetic field lines of the assumed NS dipolar field toward 
one or both of the NS magnetic poles ($B \geq 10^{12}$\ Gauss).
This plasma falls onto the NS surface at a considerable 
fraction (up to $\sim$0.5) of the speed of light, giving up 
its kinetic energy before it merges with the NS surface.
Observed X-ray luminosities for these pulsars range 
from $10^{34-35}$ ergs s$^{-1}$ for \xper\ \citep{dbrg98}, 
to near the Eddington limit ($\geq 10^{38}$ergs s$^{-1}$) 
for \lmcxfour\ \citep{lrp+91}, to possibly 
super-Eddington ($\geq 5 \times 10^{39}$ergs s$^{-1}$) 
for ULX M82 X$-$2 \citep{bhw+14}.

The physical processes involved in stopping the
accreting plasma flow as it merges onto the NS are varied. 
In the case of low-luminosity accretion, \citet{lr82} invoked a 
collisionless shock above the NS surface to slow 
the flow to subsonic velocities. 
Gas pressure then causes the flow to settle sub-sonically 
onto the NS surface.
However, there has always been doubt about the ability of a 
shock to form in the presence 
of such a strong magnetic field.
Thus, many models have appealed to Coulomb scattering of fast protons 
in the downward accretion flow against electrons in the NS 
atmosphere \citep[e.g., see][]{bs75b,hmkg84,nsw93}.
This model suggests that in the low to moderate luminosity
regime, collisional processes decelerate the 
flow and radiation pressure is not important.
In the high luminosity regime near or 
above $10^{37}$ ergs s$^{-1}$, however, radiation pressure is 
expected to be the principal agent slowing the plasma as it 
approaches the NS surface \citep{kajh96,bks+12}.

Each of the models described above has significant drawbacks.
Only the study by \citet{mn85} was able to roughly reproduce the 
emergent X-ray spectrum of \herxone, and that was 
based on a static hot slab configuration.
Furthermore, none of these models has successfully been translated into 
a tool for extracting physical accretion flow parameters 
via the fitting of real X-ray spectral data.
This situation has changed with the emergence of the bulk and 
thermal Comptonization model of \citet[][hereafter BW]{bw07}.
BW analytically modeled the channeled steady-state accretion 
flow at the surface of the NS as a radiating plasma heated by a 
radiation-dominated shock above the NS surface.
This plasma Compton-reprocesses seed photons injected into the
column via bremsstrahlung emission, cyclotron emission 
at the cyclotron resonant frequency, and blackbody emission. 
The bremsstrahlung and cyclotron photons are injected 
throughout the column, and the blackbody photons
are injected at the surface of the thermal mound, 
located at the base of the accretion flow \citep[e.g.,][]{do73}.
The combination of thermal and bulk Comptonization naturally 
generates the characteristic cutoff power law X-ray spectra observed 
in these sources.
The radiation-dominated shock means that intense radiation (rather than
a collisionless shock plus gas pressure or Coulomb scattering of
accreting protons) provides the deceleration of the accreting flow.

Three related but distinct motivations exist for the present work.
First, most spectral studies of accreting X-ray pulsars utilize 
phenomenological models such as cutoff power laws that are simple, 
easy to compute, and do a reasonable job in 
describing the real broad-band continua of accreting X-ray pulsars.
However, this agreement comes at a price, namely, that different
expressions for power law continua have been used to fit different 
pulsars, and the resulting derived model parameters do not lend 
themselves to cross-comparison.
Second, even if the fits obtained are statistically satisfactory, 
they yield almost no information about the physical parameters of the 
accretion flows.
For example, one cannot translate the value of the power law slope 
into meaningful values for the physical parameters of the accretion flow.
Finally, the derived fit parameters such as the cyclotron line 
centroid energy may depend on the adopted continuum model. 
This is important because some recent studies have shown that in 
several sources, using the broad functional continuum fits, the resulting 
centroid energies of the cyclotron lines which give the magnetic field 
strength can vary with luminosity \citep[e.g., see][]{ssp+07,tls10,vks11}.
Indeed, for 4U\,0115$+$63, \citet{mfk+13} and \citet{btl13} 
found that whether or not the cyclotron line centroid energy 
changed with luminosity depended on what functional 
model was invoked to represent the X-ray spectral continuum. 
Thus a critical question is, how is the uncertainty in the modeling
of the spectral continuum around the cyclotron line affecting the value 
of the field strength derived from the fits to the cyclotron lines?
This question cannot be resolved unless one utilizes a meaningful
physical model for the continuum, such as the BW formulation.

The BW model for the spectral formation in accreting X-ray pulsars 
can be tested against high signal-to-noise data sets from high 
luminosity pulsars.
The Nuclear Spectroscopic Telescope Array \citep[NuSTAR, ][]{hcc+13} 
satellite is a new resource of high-quality data that are ideal
for this purpose, both because it covers a large range in X-ray 
energy (3$-$80 keV)
and because the use of X-ray imaging gives it a low background 
and high signal-to-noise across the energy range.  
The NuSTAR spectral data we study in this paper have already 
been presented in a previous paper \citep{fgs+13}.
Our purpose here is to apply the BW spectral model in a manner 
consistent with previous spectral studies, and to show that this new 
model successfully reproduces the phase-averaged X-ray spectrum 
for the prominent accreting X-ray pulsar source \herxone\ over 
the entire NuSTAR energy range. 
This will also yield the highest-quality determination of the 
physical source parameters, such as column radius, 
electron temperature, etc.

\section{Model Implementation}
\label{sec:analytics}

The model we implement assumes a cylindrically collimated 
radiation-dominated radiative shock (RDRS) in the accretion 
flow confined by the NS magnetic field. 
By radiation-dominated we mean that the total pressure is 
overwhelmingly radiation pressure and the gas pressure is 
negligible. 
We envision an accretion flow entering the top of the cylindrical 
funnel at roughly free-fall velocity, passing through a sonic 
surface we identify as a radiation-dominated shock ``front" that
is a few electron scattering lengths thick, and then
coming to a halt just below the thermal mound, located just 
above the NS surface.
As the flow transitions through the thermal mound, it merges with 
the NS interior. 
We do not allow for the dipolar spreading of the field lines 
with altitude above the NS surface.

The radiation transport solution we apply is based on the 
formalism developed by BW, who derived an analytical solution 
to the time-independent 
cylindrical plane-parallel transport equation 
including the Kompaneets term for the photon distribution 
function $f(z,\epsilon)$ in the accretion column:
\begin{equation}
v \frac{\partial f}{\partial z} =
\frac{d v}{d z} \frac{\epsilon}{3} \frac{\partial f}{\partial \epsilon}
+ \frac{\partial}{\partial z} \Bigl( \frac{c}{3 n_e \sigma_{\parallel}} \frac{\partial f}{\partial z}  \Bigr) -\frac{f}{t_{\rm esc}} 
+ \frac{n_e \bar{\sigma} c}{m_e c^2} \frac{1}{\epsilon^2} \frac{\partial}{\partial \epsilon} 
\Bigl[ \epsilon^4 \Bigl( f + k T_e \frac{\partial f}{\partial \epsilon}  \Bigr) \Bigr] + \frac{Q(z,\epsilon)}{\pi r_0^2} \, .
\label{eq:radtransport}
\end{equation}
Here, $z$ is the upward distance from the stellar surface 
along the columnar axis, $v < 0$ is the inflow velocity 
(negative values denote downward velocities), $\epsilon$ 
is the photon energy, $T_e$ is the electron temperature of the
column plasma, $r_0$ is the column radius, 
$n_e$ is the electron number density, 
$\sigma_{\parallel}$ is the electron scattering 
cross section parallel to the magnetic field, $\bar{\sigma}$
is the angle-averaged mean scattering cross section
that regulates the Compton scattering process, and the 
other symbols have their usual meanings.
$Q(z,\epsilon)$ denotes the source of seed photons for the 
Comptonization process and $t_{\rm esc}$ is the time scale on 
which the scattered photons escape from the sides of the column.
The source function $Q(z,\epsilon)$ includes three principal 
emission mechanisms that cool the plasma: 
bremsstrahlung from the entire plasma volume between the
sonic point and the NS surface, cyclotron emission 
from throughout the column, but only at
the cyclotron resonant energy, and finally
blackbody emission from a dense thermal mound
at the base of the flow, where the accreted plasma merges
with the NS.
The BW solution method was to first  
obtain the Green's function $f_{\rm G}(z_0,z,\epsilon_0,\epsilon)$ that gives the 
radiation distribution at altitude $z$ and photon energy $\epsilon$ 
resulting from the injection of photons at height $z_0$ 
and energy $\epsilon_0$.
In order to obtain $f_{\rm G}$, equation (1) must be made 
separable by substituting a simple linear velocity profile 
in terms of the flow optical depth along 
the column ($\tau$).
This is accomplished by setting 
\begin{eqnarray}
v(\tau) = - \alpha c \tau \ ,
\label{eq:velocitylaw}
\end{eqnarray}
where $\alpha$ is a constant defined below and is of order 
unity \citep[see][]{ls82}.
The transport equation can now be solved as separate 
spatial and energy differential equations.
See BW for details of the analytic treatment.

In the expressions we will evaluate we will need the 
similarity variables $\delta$, $\xi$, 
and $\alpha$ \citep[see also ][]{fbs+09}, and also $t_{\rm esc}$.
These are related to the input physical parameters via 
the expressions (BW),  
\begin{equation}
\delta \, = \, \frac{\alpha}{3} \frac{\sigma_{\parallel}}{\bar{\sigma}} \frac{m_e c^2}{k T_e} \, = \, 4 \frac{y_{\mathrm{bulk}}}{y_{\mathrm{thermal}}},
\label{eq:delta}
\end{equation}
\begin{equation}
\xi \, = \, \frac{\pi r_0 m_p c}{\dot{M} \sqrt{\sigma_{\perp} \sigma_{\parallel}}},
\label{eq:xi}
\end{equation}
\begin{equation}
\alpha \, = \, \frac{32 \sqrt{3}}{49 \, {\rm ln}(7/3)} \frac{G M_* \xi}{R_* c^2},
\label{eq:alpha}
\end{equation}
and,
\begin{equation}
t_{\rm esc}(z) \, = \, \frac{\dot{M} \sigma_{\perp}}{\pi m_p c |v| }.
\end{equation}
Here, $\dot{M}$ is the mass accretion rate, $M_*$ and $R_*$ are the 
mass and radius of the neutron star, respectively, $\sigma_{\perp}$ is 
the scattering cross section perpendicular to the magnetic field, 
and $y_{\mathrm{bulk}}$ and $y_{\mathrm{thermal}}$ are the 
respective Comptonization $y$-parameters defined in \citet{rl79}.

``Bulk" or ``dynamical" Comptonization in this context is the 
transfer of energy from the protons in the flow to the electrons
via Coulomb coupling, and then to the seed 
photons via first-order Fermi energization in the presence 
of the compressing flow through the shock front.
``Thermal" Comptonization in the BW model is the process whereby
thermal electrons in the background plasma scatter off of
the injected seed photons, giving up energy to those photons,
and with those photons ultimately escaping out of the sides of the column.
Thus, the parameter $\delta$ will help us to understand the relative 
importance of bulk and thermal Comptonization in the flow models 
we obtain for each source.

Given the analytical solution for the Green's function, 
the problem is reduced to specifying 
the source term $Q(z,\epsilon)$ for each of the 
three physical emission processes that supply the seed photons.
We approximate the monochromatic source term for cyclotron photon 
production according the prescription of \citet{akl87}.
Comptonized cyclotron emission is then computed according to 
equation (117) in BW.
In our model, cyclotron emission injects photons only at the 
cyclotron energy in the plasma ($\epsilon_c$), but also 
continuously in  height ($z$) from the radiating region 
between the sonic surface and the thermal mound.
We compute the contribution to the total column-integrated 
spectrum at energy $\epsilon$ by Comptonized cyclotron 
emission using the series expansion
\begin{eqnarray}
\Phi_{\epsilon}^{\rm cyc}(\epsilon) & = &
\frac{3.43 \times 10^{-16} \dot{M} \,\, H\bigl(\epsilon_c/(k T_e)\bigr) \,\, \xi^2 \sqrt{\alpha^3 w} \epsilon^{\kappa - 2}}{\bar{\sigma} \epsilon_c^{\kappa + 3/2} \exp[(\epsilon_c + \epsilon)/(2 k T_e)]}
\sum_{n = 0}^{\infty} \frac{\Gamma(\mu_n-\kappa+1/2) n!}{\Gamma(1+2\mu_n)\Gamma(n+1/2)}  X_n A_n \nonumber \\
& & \times \,\, M_{\kappa,\mu_n}\Bigl( {\rm min}[\frac{\epsilon}{kT_e},\frac{\epsilon_c}{kT_e} ] \Bigr) 
W_{\kappa,\mu_n}\Bigl( {\rm max}[\frac{\epsilon}{kT_e},\frac{\epsilon_c}{kT_e} ] \Bigr) , 
\end{eqnarray}
where $\epsilon_c= e h B/(2 \pi m_e c)$ is the 
cyclotron energy, $w = \sqrt{9+12 \xi^2}$, $B$ is the 
magnetic field strength, and the sum is over the 
eigenvalue index $n$.
The indices $\kappa$ and $\mu_n$ are defined as 
\begin{equation}
\kappa = \frac{1}{2}(\delta+4), \,\, {\rm and,} \,\,\,\, \mu_n = 
\frac{1}{2}(3-\delta)^2+\delta [(4n+1) (9 + 12 \xi^2)^{1/2} + 3],
\end{equation}
and the function $H\bigl(\epsilon_c/(k T_e)\bigr)$ is 
defined in \citet{akl87} as
\begin{equation}
H \biggl( {\frac{\epsilon_c}{k T_e}} \biggr) = 
 \begin{cases} 
  0.41 &\mbox{$\epsilon_c/k T_e \geq 7.5$},\\
  0.15\sqrt{\epsilon_c/k T_e} & \mbox{$\epsilon_c/k T_e < 7.5$}.
 \end{cases}
\end{equation}
$M_{\kappa,\mu_n}$ and $W_{\kappa,\mu_n}$ are Whittaker 
functions \citep{as70}.
The functions $A_n$ and $X_n$ are analytic, given in BW, and 
can be evaluated from input parameters.

Blackbody seed photons are emitted only at the height
of the thermal mound surface, but continuously in energy.
The blackbody contribution to the total column-integrated 
spectrum at energy $\epsilon$  
can be expressed by the series expansion
\begin{eqnarray}
\Phi_{\epsilon}^{\rm bb}(\epsilon) & = & \frac{6 \pi^2 r_0^2 \delta \xi^2 k T_e \sqrt{2 \alpha^3 w} \exp(\frac{3 \alpha \tau_{\rm th}^2}{2}) 
\epsilon^{\kappa-2} \exp(- \frac{\epsilon}{2kT_e}) }{c^2 h^3} \,\,\,\,\, \nonumber \\
& & \times \,\, \sum_{n = 0}^{\infty} \frac{\Gamma(\mu_n-\kappa+1/2) n! X_n g_n(\tau_{\rm th})}{\Gamma(1+2\mu_n)\Gamma(n+1/2)} \,\,\,\,\, 
\Biggl[ W_{\kappa,\mu_n}\Bigl( \frac{\epsilon}{kT_e} \Bigr) \int_0^{\epsilon} M_{\kappa,\mu_n} \Bigl( \frac{\epsilon_0}{kT_e} \Bigr)
\frac{\epsilon_0^{2-\kappa} \exp(\frac{\epsilon_0}{2kT_e})}{\exp(\frac{\epsilon_0}{kT_{\rm th}})-1} d \epsilon_0 \nonumber \\
& & + \, M_{\kappa,\mu_n}\Bigl( \frac{\epsilon}{kT_e} \Bigr) \int_{\epsilon}^{\infty} W_{\kappa,\mu_n} \Bigl( \frac{\epsilon_0}{kT_e} \Bigr)
\frac{\epsilon_0^{2-\kappa} \exp(\frac{\epsilon_0}{2kT_e})}{\exp(\frac{\epsilon_0}{kT_{\rm th}})-1} d \epsilon_0 \Biggr] ,
\end{eqnarray}
where the $g_n$ functions are the Laguerre 
polynomials, $\tau_{\rm th}$ is the optical
depth at the top of the thermal mound, given by
\begin{equation}
\tau_{\rm th} \, = \, 2.64\times10^{28}\frac{\dot{M} R_*}{M_* r_0^{3/2} T_{\rm th}^{7/4} \xi },
\end{equation}
and $T_{\rm th}$ is the temperature of the thermal mound, given
by $T_{\rm th} = 2.32\times10^3 {\dot{M}}^{2/5} r_0^{-2/3}$ 
in cgs units.
The two integrals in Equation (10) must be evaluated numerically.
We note that the second of these integrals can take a 
significant amount of time to compute.
We utilize a Gaussian-Legendre Quadrature integration scheme to 
handle the trade-off between increased computational speed and 
maintaining sufficient numerical accuracy.
When the blackbody integrals are excluded, by setting the 
software switch appropriately (see Section~\ref{sec:results}), 
the calculation currently speeds up by roughly a factor of ten.

Bremsstrahlung seed photons will be emitted at all heights 
within the hot plasma and continuously across the 
X-ray energy range. 
Bremsstrahlung emission is computed according to 
equations (128) and (129) 
in BW.
The full column-integrated spectral flux at energy $\epsilon$ 
is given in this case by the series expansion
\begin{eqnarray}
\Phi_{\epsilon}^{\rm ff}(\epsilon) & = &
\frac{2.80 \times 10^{-12} \dot{M} \xi^2 \sqrt{\alpha^3 w} \epsilon^{\kappa - 2} e^{-\epsilon/(2 k T_e)}}{\bar{\sigma} (k T_e)^{\kappa + 3/2}}
\sum_{n = 0}^{\infty} \frac{\Gamma(\mu_n-\kappa+1/2) n!}{\Gamma(1+2\mu_n)\Gamma(n+1/2)}  X_n A_n \\
& & \times \,\, \int_{\chi_{\rm abs}}^{\infty} (\frac{\epsilon_0}{kT_e})^{-1-\kappa} \exp(-\frac{\epsilon_0}{2kT_e}) M_{\kappa,\mu_n}\Bigl( {\rm min}[\frac{\epsilon}{kT_e},\frac{\epsilon_0}{kT_e} ] \Bigr) 
W_{\kappa,\mu_n}\Bigl( {\rm max}[\frac{\epsilon}{kT_e},\frac{\epsilon_0}{kT_e} ] \Bigr) d\epsilon_0 , \nonumber
\end{eqnarray}
where the integration is over the full input spectrum of the 
bremsstrahlung emission in the plasma from the self-absorption
cut-off energy to infinity.
The dimensionless self-absorption cutoff energy ($\chi_{\rm abs}$) is based on 
equation (127) of BW.
The density factor is the geometric 
mean of the density at the upper surface of the
radiating region [see Equation (80) of BW] and at the thermal mound.

Once the three individual components of the spectral flux density are 
computed, and assuming the problem is linear (see BW), 
we can add the three components together to obtain the full X-ray 
photon spectral continuum for the accreting pulsar:
\begin{equation}
F_{\epsilon} (\epsilon) \equiv \frac{ \Phi_{\epsilon}^{\rm cyc} (\epsilon) + 
\Phi_{\epsilon}^{\rm bb} (\epsilon) + 
\Phi_{\epsilon}^{\rm ff} (\epsilon) }{4 \pi D^2} ,
\end{equation}
where $D$ is the distance to the source and is an additional 
model input parameter. 
The function $F_{\epsilon}(\epsilon)$ gives the spectral photon flux at energy $\epsilon$ from the
Comptonization of the bremsstrahlung, cyclotron and thermal mound blackbody
seed photons in units of photons cm$^{-2}$ s$^{-1}$ keV$^{-1}$.

The BW model assumes that the post-shock plasma 
electron temperature ($T_e$) is constant in
the region between the sonic point and the thermal mound. 
Hence the structural model
includes no thermodynamic feedback
between the various emission components that ultimately cool
the plasma as it approaches the NS surface and the 
radiation-hydrodynamic structure of the decelerating 
shocked plasma. 
Moreover, because the velocity law in the BW model is assumed to be a linear
relation given by Equation (2), the resulting structure of the 
accretion column between the sonic surface of the radiative shock 
and the NS surface will not necessarily 
conform to how nature would actually decelerate and stop a real 
accretion flow in this environment. 
Consequently, the analytical model of BW will not automatically conserve 
energy when applied to a specific source, in the sense that the 
resulting emergent energy-integrated luminosity $L_{\rm X}$ may not equal the 
accretion luminosity, $L_{\rm acc}= G \dot{M} M_*/R_*$. 
In our modeling approach, we enforce the energy conservation requirement 
$L_{\rm X}= L_{\rm acc}$ as part of the fitting 
procedure, and this constrains the accretion rate $\dot M$, 
as further discussed below. 
However, as part of our investigation of model diagnostics, 
we will allow $\dot M$ to vary by $\pm 30$\% around its best-fitted value
in order to explore the sensitivity of the 
model parameters to accretion rate variations in a context where
the energy conservation requirement is relaxed.

\section{X-ray Spectrum of \herxone\ in NuSTAR}
\label{sec:herx1nustarobs}

In order to compare our theoretical RDRS model directly with 
an observed accreting X-ray pulsar spectrum, we would like to 
study a high-luminosity pulsar 
with as little intervening absorbing material modifying the
observed spectrum as possible. 
\herxone\ is an accreting X-ray pulsar with a 1.24 s spin period 
and a 1.7 day binary orbital period and it has an observed 
luminosity near $\sim 4\times10^{37}$ ergs s$^{-1}$.  
The interstellar column density to the \herxone\ system is 
relatively low (see below), but a concern for \herxone\ is 
that the NS is believed to be surrounded by a precessing accretion 
disk that sometimes intervenes between the observer and 
the NS X-ray source, bringing about a 35-day super-orbital 
cycle \citep[e.g., see][]{sl99}. 
However, there are intervals during the accretion disk 35-day 
cycle when the disk precesses out of our line of sight and 
we have a relatively unobstructed view of the spin-phase 
averaged ``main-on'' X-ray spectrum produced by the gas accreting onto 
the surface of the NS. 
The spectrum we fit below is taken from the ``main-on" 
section of the 35-day super-orbital period and is the 
same section as the ``II" section from \citet{fgs+13}.

Another concern for \herxone\ is that in order for our 
comparison to be fully valid we need a system whose 
luminosity puts it in either the critical or the moderately 
sub-critical luminosity range of \citet{bks+12}. 
In \citeauthor{bks+12} model, the entire deceleration of the
flow is accomplished by radiation pressure when the luminosity
is above the critical luminosity $L_{\rm crit}$. 
If the luminosity is below this limit but still relatively 
high, the flow deceleration is accomplished mostly by 
radiation pressure with a layer of Coulomb collisional 
deceleration near the neutron star surface. 
In either case, the flow will be radiation-dominated 
in the spectral formation region below the actual shock front. 
During our observation \herxone\ was radiating with 
luminosity $\sim4.9\times10^{37}$ ergs s$^{-1}$ (see below) 
which just places it in the sub-critical range according to 
\citeauthor{bks+12} and in the super-critical 
range according to \citet{mstp15}. 
In deriving the boundary between the super-critical and sub-critical 
ranges, \citeauthor{bks+12} found that the critical luminosity, 
$L_{\rm crit} \propto \Lambda^{-7/5}$ where $\Lambda$ is a 
geometrical parameter characterizing whether the NS accretes 
from a wind or from a disk. 
For disk accretion a value $\Lambda \sim 0.5$ may be more 
appropriate than the value $\Lambda \sim 0.1$ assumed by 
\citeauthor{bks+12}
Using this value for $\Lambda \sim 0.5$ in \citeauthor{bks+12}'s 
equation (55) yields $L_{\rm crit} \sim 7.3\times10^{36}$ ergs s$^{-1}$ 
which is significantly below the \herxone\ luminosity that we observe. 
Thus, we conclude here that \herxone\ is most likely in, 
or at least very close to, the super-critical accretion 
rate range as obtained by both \citet{bks+12} and \citet{mstp15}
and our model is applicable.

The X-ray spectrum of \herxone\ was extracted  
using \textit{nupipeline} v.1.3.1 
with standard filtering (e.g., for SAA passages and Earth occultations) 
applied to the data from from 
NuSTAR \citep{hcc+13} ObsID 30002006005. 
This observation occurred on 22 September 2012 from 04:20:32 
to 18:35:00 UTC and the effective exposures for the two focal 
plane modules, FPMA and FPMB, are 21.9 ks and 22.1 ks, respectively.
From the cleaned event files we extracted the spectrum 
using \textit{nuproducts} from a 120 arc second radius region 
centered on the J2000 coordinates of \herxone.
The background spectrum was extracted from an 87 arc second 
radius region to the south of the source. 

We fitted the \herxone\ NuSTAR spectrum between 4 and 79 keV. 
When we tried to fit the spectrum below 4 keV an apparent soft 
excess forced us to add a thermal component to the fit having
a very poorly constrained temperature and normalization.
So, we restricted the fitted energies to 4 keV and above.
We did not apply any systematic error in our fit to the RDRS model.
Our source spectrum is everywhere larger than the 
background spectrum except above 60 keV where the
two become comparable. 
We implemented a channel grouping using the FTOOLS command
{\it grppha}.
From low to high energies we progressively included more 
of the raw spectral channels into each final spectral bin so 
that the final fitted spectrum had a total of 512 bins from 
both detector modules across the 4 to 79 keV range.
This balances the need to reduce the number of nonindependent 
spectral channels and provide a sufficiently high signal-to-noise 
with the need to maintain sufficient resolution in the final 
spectrum to be able to constrain real energy-dependent 
features in the analysis. 
Finally we note that all errors are reported at the 90\% 
confidence level in this paper.

\begin{figure}
\begin{center}
\includegraphics[width=5.67in]{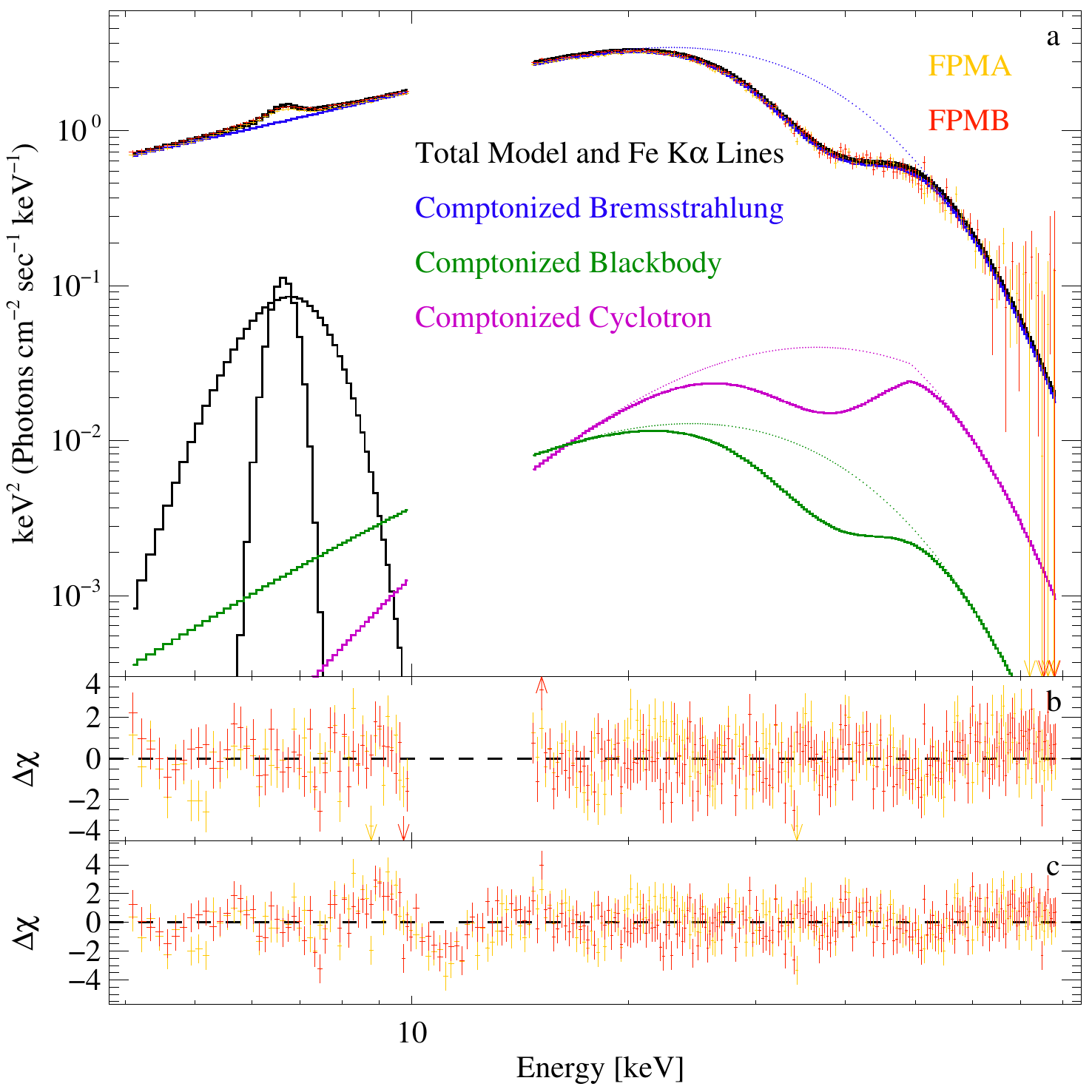}
\hfill
\caption{\footnotesize 
(a) \herxone\ 4$-$79 keV X-ray 
unfolded (after \citet{nhtm+11}) spectrum 
from NuSTAR with residuals shown in the lower two panels.
The two NuSTAR focal plane detectors are shown and 
the spectrum has been rebinned for clarity (see text). 
The final reduced $\chi^2$ is 1.212 for 450 degrees of freedom.
The individual Comptonized emission components
are shown.
The dotted lines represent the Comptonized components with the
multiplicative effects of cyclotron absorption removed.
The model is dominated by Comptonized bremsstrahlung emission
as suggested by BW.
Close examination of the Comptonized cyclotron
component reveals a mild discontinuity that 
results from the monochromatic injection of
cyclotron photons at the local cyclotron 
energy at the neutron star surface (49.3 keV).
(b) The top set of residuals are from our best 
fit [panel (a)] with the
10$-$14.5 keV region of the spectrum removed (see text)
and all the two Fe emission line parameters free to vary.
(c) The bottom residuals are from our best fit to the full
energy range 4$-$79 keV and with the Fe-line widths fixed
at the values obtained by \citet{fgs+13}. 
The small systematic deviations in the 
9$-$14 keV range are most likely due
to uncalibrated effects from the Tungsten 
coating on the NuSTAR mirrors.}\label{fig:herx1spec}
\end{center}
\end{figure}

\section{Results}
\label{sec:results}

We performed our spectral fit in the XSPEC v12.8.2 and
12.9.0 \citep{arna96} 
spectral analysis environments.
We verified that the resulting fits are 
consistent across both versions of XSPEC.
23 parameters describe the spectral continuum and 
emission lines that NuSTAR observes in the 4 to 79 keV range.
Six parameters describe two iron emission lines 
(we use the {\it gaussian} function for both lines), three
describe the cyclotron resonant scattering absorption 
feature (CRSF), one parameter
describes the interstellar absorption ($N_{\mathrm{H}}$), 
and one parameter describes the cross-normalization of 
the FPMA and FPMB modules (C$_{\rm FPM}$).
The absorbing column is frozen at 
$N_{\mathrm{H}}\,=\,1.7 \times 10^{20}$ cm$^{-2}$, 
a value that approximates the galactic absorption 
to \herxone\ \citep{fgs+13}.
We utilized the \textit{tbabs} absorption model with the ``WILM" 
abundance table and the ``BCMC" cross sections \citep{fam00}.
This leaves 12 parameters to describe the RDRS model. 
One of these is a RDRS model normalization 
that we freeze at 1.0.
One is a numerical switch that allows us to turn on 
or off any of the three principal seed photon processes for 
the Comptonization (see Section~\ref{sec:analytics}).
We freeze the distance to the \herxone\ system at $D = 6.6$\ kpc
\citep{rqs+97}, and the NS mass and radius are set to their
standard values ($1.4 \,\, M_{\odot}$ and $10$ km).
The distance to \herxone\ obtained by \citeauthor{rqs+97} 
of $6.6_{-0.4}^{+0.4}$\ kpc is sufficiently 
consistent, for our purposes, 
with a more recently determined distance of $6.1_{-0.4}^{+1.0}$ kpc
obtained by \citet{la14}.
Thus, we adopt the \citeauthor{rqs+97} distance for these calculations.
The value of the input magnetic field strength to the RDRS model
is tied to the centroid of the fitted cyclotron absorption 
line ($E_{\mathrm{CRSF}}$) via 
$B/(10^{12} \mathrm{Gauss}) \, = B_{12} \, = \, (1+z_*) E_{\mathrm{CRSF}} /11.57 $ where $z_*$ 
is the gravitational redshift to the NS surface. 
The six remaining parameters describing the RDRS 
model (BW) are: 
mass accretion rate ($\dot{M}$),
plasma Comptonizing temperature ($T_e$),
accretion cap radius ($r_0$),
scattering cross section perpendicular to the magnetic 
field ($\sigma_{\perp}$),
scattering cross section parallel to the magnetic 
field ($\sigma_{\parallel}$),
and average scattering cross section ($\bar{\sigma}$).
The scattering cross section perpendicular to the 
magnetic field ($\sigma_{\perp}$) is frozen at the 
Thomson electron scattering cross section ($\sigma_{\mathrm{T}}$).
This leaves five free parameters, $\dot M$, $T_e$, $r_0$, 
$\sigma_{\parallel}$, and $\bar{\sigma}$, to
describe the X-ray continuum.

The fit to the observed NuSTAR spectrum of \herxone\ is shown 
in Figure~\ref{fig:herx1spec}.
Initially, we tried to fit the full energy range of 4 to 79 keV 
to the calculated radiation-dominated shock spectrum. 
However, in the resulting fit, the broad iron line 
component (FeK$\alpha_{\rm b}$) 
significantly modifies the continuum in the range below 15 keV. 
This broad iron feature apparently smooths deviations of the 
observed spectrum from the computed model in the 8$-$15 keV range.
A similar difficulty was encountered by \citet{fgs+13} in
this same spectral region. 
To test for this possibility, we fixed the two iron line widths 
at the values found by \citet{fgs+13}. 
As a result of this, just as found by \citeauthor{fgs+13}, the observed 
spectrum in the energy range of 9$-$15 keV has systematic residuals 
in comparison with the computed model spectrum. 
The reduced $\chi^2$ value is 1.42 for 500 degrees of freedom 
and the model residuals are shown in the bottom panel in 
Figure~\ref{fig:herx1spec}. 
These systematic residuals are most likely caused by 
imperfections in the response matrices due to Tungsten the 
edges from 8.39 to 12.10 keV induced by the X-ray mirror 
coatings \citep{mhm+15}. 
Following the procedure of \citet{fgs+13}, we removed 
the spectral points in the energy range 10$-$14.5 keV. 
The fit is now significantly better with reduced $\chi^2$ 
of 1.212 for 450 degrees of freedom which is as good as any 
of the empirical fits obtained by \citet{fgs+13}. 
The best-fit parameters are given in Table 1
along with their formal 90\% confidence errors. 
The continuum fit with the radiation-dominated shock model 
conforms very well to the observed background subtracted 
Her X- 1 spectrum. 
Examination of the contributions to $\chi^2$
as a function of energy suggest that there are still some 
deviations at the highest energies in our spectral range 
(65$-$79 keV). 
This is most likely because the background spectrum becomes 
as large or larger than the \herxone\ source spectrum above 
about 65 keV.

\subsection{Accretion Rate}
\label{subsec:accrate}

The accretion rate is ultimately determined by requiring that the total 
luminosity $L_{\rm X}$ in the three Comptonized spectral 
components (bremsstrahlung, cyclotron, and blackbody emission) 
integrated over the range 0.1-100 keV is equal to the \herxone\ 
accretion luminosity ($L_{\mathrm{acc}} \, = \, G M_* \dot{M}/R_*$). 
We also require that the observed flux in the 5$-$75 keV band 
equals the model flux in this same energy band.
This leads to the need to iterate on the mass accretion 
rate ($\dot{M}$) by first estimating the accretion 
rate based on an approximate luminosity and then refining 
that estimate as one fits a series of models to the data until 
both of these conditions are satisfied 
for one set of model parameters. 
One source of uncertainty in the accretion rate is the 
uncertainty in the distance to \herxone. 
There is also the possibility of some thermal energy being 
transported into the star at the base of the column 
\citep[e.g.,][]{bs76}. 
If energy is transported into the star, then it would 
obviously create an offset between the values of $L_{\rm acc}$ 
and $L_{\rm X}$. 
In the absence of a comprehensive model that treats 
both the stellar surface and the shocked emitting region, 
it is not possible to specify the exact surface 
boundary condition. 
BW assumed that the radiated energy vanishes at 
the stellar surface, and we adopt the BW surface boundary condition
in our work, by setting $L_{\rm X} = L_{\mathrm{acc}} = G M_* \dot{M}/R_*$,
where the value of $L_{\rm X}$ is set equal to the luminosity 
emitted in the energy range 0.1$-$100 keV. 
In Section~\ref{subsec:modlimuncer}, we discuss the 
results obtained for the model parameters by artificially 
varying $\dot M$ around the value computed by 
setting $L_{\rm X} = L_{\rm acc}$.

\subsection{Model Parameters}
\label{subsec:modparams}

In general, the parameters from our fit are not too different
from the approximate parameters obtained by BW for \herxone.
BW found the spectral shape was approximated sufficiently
well in their low resolution limit by $r_0 \, = \, 44$\ m whereas 
we obtain $r_0 \, = \, 107.0^{+1.7}_{-1.8}$ m, and BW found the
column to be somewhat hotter at $kT_e \sim 5.4$ keV, compared to
our fitted $kT_e \sim 4.58^{+0.07}_{-0.08}$ keV.
The temperature we find for the thermal mound in our model 
is only a little below this at 3.8 keV.
This temperature is about twice the Eddington temperature for the 
NS surface of $\sim 1.9$ keV.
This is reasonable because blackbody photons are Compton scattered 
in the hot plasma, re-directing them out the sides of the column. 
This reduces the ability of those photons to halt the accretion
flow by exchanging outward directed momentum with the 
falling electrons.
Finally, note that this thermal mound temperature is significantly 
lower than the effective temperature of $\sim 9.1$ keV predicted 
by \citet{mstp15} in the disk accretion case for our 
model parameters with $\Lambda \sim 0.5$.

At an estimated distance of 6.6 kpc \citep{rqs+97}, based on
the fitted model parameters, for this observation the 0.1$-$100 keV  
luminosity of \herxone\ is about
$4.9 \times 10^{37}$ ergs s$^{-1}$ and we expect this 
accretion flow to be radiation-dominated near the NS surface.
As found by BW, Comptonized bremsstrahlung photons
constitute the bulk of the observed emission in hard
X-rays and Comptonized cyclotron and thermal mound blackbody photons 
are not significant contributors.
Our model fit conforms to the expectation (see BW) for the resulting 
mean scattering cross sections in that we do indeed find that  
$\sigma_{\parallel} \, < \, \bar{\sigma} \, < \, \sigma_{\perp}$. 
When we compare the flux we derive from the fit in the 5$-$60 keV 
energy range with that of \citet{fgs+13}, we find after accounting 
for the new calibration released in October 
2013\footnote{http://heasarc.gsfc.nasa.gov/docs/heasarc/caldb/nustar/docs/release\_20131007.txt.},
our 5$-$60 keV flux agrees with the \citeauthor{fgs+13} flux
within 2\%.
When we compare our derived value of $\xi \sim 1.355$ with 
the value $\xi$ should have in a pure radiation-dominated shock ($\xi \sim 1.15$)
(see BW) we are about 18\% high which we believe is an indication
of the divergence of our assumed ``hot slab" structure with
approximate velocity law to a full radiation-hydrodynamic solution
in which the plasma comes to rest at the NS surface. 
Comparing our result to the approximate result of BW, we obtain
$\delta \sim 2.38$, which is still a moderate value. 
For this $\delta$-value 
equation (3) yields $y_{\rm bulk} \sim  0.59 \,\, y_{\rm thermal}$
which indicates that in our formulation, {\it bulk} Comptonization
is slightly less important than {\it thermal} Comptonization 
in determining the overall \herxone\ X-ray spectral shape. 
That thermal Comptonization is important is shown by the fact that
the \herxone\ spectrum is relatively flat at intermediate energies 
and turns over above 30 keV.
However, thermal Comptonization is not so strong (i.e., saturated) 
in our model that a Wien peak is formed in the spectrum.

\subsection{Iron Emission}
\label{subsec:ironemis}

We model the iron line complex as consisting of a narrow line
at $\sim$6.6 keV (the ``n" subscript) and a broad line at the 
lower energy of $\sim$6.5 keV (the ``b" subscript).
This is similar to the iron line model of \citet{fgs+13}
We have also removed the range 10$-$14.5 keV as did \citeauthor{fgs+13}
As we noted above retaining this region would have the effect 
of broadening one of the Fe-lines to ``smooth over" some of 
the calibration residuals in this energy range so we delete 
this section of the energy spectrum.
Furthermore, we only fit the NuSTAR X-ray spectrum here while
\citeauthor{fgs+13} fit the combined spectra from NuSTAR and
Suzaku in the region below 10 keV and Suzaku had higher 
spectral resolution than NuSTAR.
\citeauthor{fgs+13} fit a number of different continuum models and
we compare our fit to their ``HighE" cutoff power law model from 
their Table 3.
For the ``narrow" iron line, \citet{fgs+13} found an energy 
centroid of $E({\rm FeK}\alpha_{\rm n}) = 6.601^{+0.017}_{-0.016}$ keV
and $\sigma({\rm FeK}\alpha_{\rm n}) = 0.25^{+0.04}_{-0.04}$ keV
which agree within the errors of the values 
we find for $E({\rm FeK}\alpha_{\rm n}) = 6.61^{+0.02}_{-0.02}$ keV
and $\sigma({\rm FeK}\alpha_{\rm n}) = 0.26^{+0.03}_{-0.04}$ keV.
Thus, the narrow Fe-line fits are essentially consistent within 
the uncertainties.
For the ``broad" iron line we obtain an energy centroid of 
$E({\rm FeK}\alpha_{\rm b}) = 6.53^{+0.07}_{-0.08}$ keV 
and $\sigma({\rm FeK}\alpha_{\rm b}) = 0.90^{+0.21}_{-0.14}$ keV
whereas \citeauthor{fgs+13} 
find $E({\rm FeK}\alpha_{\rm b}) = 6.55^{+0.05}_{-0.05}$ keV
and $\sigma({\rm FeK}\alpha_{\rm b}) = 0.82^{+0.13}_{-0.10}$ keV.
Again, our broad line agrees within the errors with the line
parameters found by \citeauthor{fgs+13}
\citet{aeiy+14} found in their analysis of Suzaku data alone
from \herxone\ that their spectral fits required a broad line in 
the 4$-$9 keV energy range.
They suggested a number of possible physical origins for this
feature but their data was not sufficient to attribute the
line emission to one unique mechanism. 
The iron line signatures in the 4$-$9 keV range of the spectrum
is complex with perhaps more than one narrower line
sitting on top of a broad continuum consisting of many
line components that are not yet full resolved by either of these
instruments. 

\subsection{Scattering Cross Sections and Cyclotron Absorption}
\label{subsec:scatcrseccycabs}

We included a CRSF absorption 
line with the \textit{gabs} multiplicative model component 
in our broad band fit.
The energy centroid of our fit is $E_{\mathrm{CRSF}} = 37.7_{-0.2}^{+0.2}$ keV, 
which falls within the error range of the value obtained by \citet{fgs+13},
$E_{\mathrm{CRSF}} = 37.40_{-0.24}^{+0.25}$ keV.
However, our line width, $\sigma_{\mathrm{CRSF}} = 7.1_{-0.2}^{+0.2}$ keV,
is somewhat larger than the \citeauthor{fgs+13} result 
of $\sigma_{\mathrm{CRSF}} = 5.76_{-0.27}^{+0.29}$ keV.
The cyclotron resonant scattering feature absorption line optical
depth can be written as $\tau_{\mathrm{CRSF}} = {\rm Strength}/(\sigma_{\mathrm{CRSF}} \sqrt{2 \pi})$.
In our fit $\tau_{\mathrm{CRSF}} \sim 0.98_{-0.06}^{+0.06}$
whereas \citeauthor{fgs+13} obtain $\tau_{\mathrm{CRSF}}\,\sim \,0.614_{-0.025}^{+0.028}$.
Thus, our CRSF absorption line is deeper than the line
found by \citeauthor{fgs+13}
This is not a large difference and may result from our
different continuum model.

BW froze the perpendicular
scattering cross section to the Thomson value ($\sigma_\perp \equiv \sigma_{\rm T}$) 
in their analytical model and we adopt this approximation here.
However, this approximation does warrant further discussion. 
Below the cyclotron resonant energy, the cross section of 
ordinary polarization mode photons (i.e., photons with electric field
vectors in the plane formed by the pulsar magnetic field and the
photon propagation direction) will be roughly constant with
energy at the Thomson value.
The extraordinary mode photons (i.e., those photons with electric 
field vectors oriented perpendicular to the plane formed by the
pulsar magnetic field and the photon propagation direction) 
have a scattering cross section significantly below Thomson at 
energies below $\epsilon_c$ because in this range the extraordinary 
mode scattering cross section varies as  
$\sigma_{\rm ext} \propto (\epsilon/\epsilon_c)^2$ 
\citep[see][]{nage81b}.
Utilizing a simple slab geometry with a numerical formulation of 
the transport equation, \citeauthor{nage81b} was able to show that at 
high Thomson scattering optical depths the emerging X-ray spectrum 
in the energy range $< 20$ keV could be dominated by extraordinary 
mode photon emission.
We are working in a parameter regime where the perpendicular 
Thomson scattering optical depth is $\tau_{\perp} \, \sim \, t_{\rm esc} c / r_0 \sim 1100$
according the BW Equation (17) and thus large.
Hence, the spectra of \herxone\ 
below the first cyclotron resonance may be dominated by 
extraordinary mode photon emission. 

Furthermore, in \herxone\ we also see that most of the emergent
radiated energy comes out near the cyclotron energy 
and thus setting the perpendicular cross section to Thomson is
not an unreasonable approximation.
We therefore model the electron scattering of both modes 
perpendicular to the magnetic field as occurring at the Thomson rate. 
Furthermore, the extraordinary mode photons interact most strongly 
with electrons at the resonance energy via absorption, which is 
followed almost immediately by the emission of a cyclotron photon 
at the same energy but in a different direction.
In the frame of the neutron star, this is essentially a resonant 
elastic scattering \citep[see][]{nage80}.
The net effect of this resonant scattering process is
negligible within the context of our model, since our 
transport equation is angle-averaged.
Therefore, the appropriate way to include this effect 
in our spectral modeling is to impose 
an absorption feature as part of the XSPEC modeling of the 
emergent spectrum, which is what we do.
More detailed future calculations of the X-ray spectra of high 
luminosity accreting X-ray pulsars are needed to explore 
this issue.

In our model, we implicitly assume that the cyclotron 
absorption feature is associated with the same magnetic 
field value as the cyclotron emission. 
However, in principle, one may consider the possibility 
of ``disconnecting'' the cyclotron emission magnetic 
field from the cyclotron absorption magnetic field, 
if one supposes that emission occurs primary in a 
separate region of the accretion column from the 
location where the cyclotron scattering feature is 
imprinted \citep[e.g., see][]{fbs+09}.
This is a further level of approximation when cast 
within the context of a cylindrical model 
that doesn't include a self-consistent calculation 
of the electron temperature or the velocity profile. 
Moreover, in the particular application to \herxone\
made in this paper, the distinction between the two 
fields will make no difference for the model fits, 
since the contribution to the observed spectrum from 
the Comptonized cyclotron component is negligible. 
In the future, the actual dipole variation of the field 
should be included, along with the flow dynamics and 
the variation of the electron temperature (see discussion 
in Section~\ref{subsec:modlimuncer}).

\subsection{Relation to Previous Work}
\label{subsec:relprevwork}

\citet{fbs+09} were the first to implement computationally 
a version of the BW formalism, with a specific 
application to the analysis of \fouruooneonefive. 
\fouruooneonefive\ has a magnetic field strength of $\sim 10^{12}$ G 
resulting in cyclotron photons being a very significant 
seed photon source for Comptonization in the accreting plasma 
near the NS surface.
Furthermore, the fundamental CRSF and as many as four harmonics 
have been detected in this source \citep{fbs+09}. 
In their modeling of the X-ray spectrum, \citeauthor{fbs+09} 
concluded that the BW Comptonization model can account for 
the X-ray spectral continuum above $\sim$9 keV as resulting 
from Comptonized cyclotron photons. 
In order to obtain a reasonable spectral fit below 9 keV, 
\citeauthor{fbs+09} found that 
an additional power law component, possibly resulting 
from Comptonized blackbody photons emerging from 
a significant fraction of the NS surface, was required.
The work of \citeauthor{fbs+09} also included an attempt to 
implement phase-resolved spectroscopy, although their 
application of the column-integrated BW model meant 
that the phase-resolved fits should really be interpreted 
as rough estimates, since they allowed the BW model 
parameters to vary as a function of the star's spin phase, 
rather than employing a single fixed model for the column, and 
allowing it to be viewed from various observation angles. 
This latter scenario, while more accurate, cannot be implemented 
without the availability of a height-dependent version of 
the BW model (rather than the column-averaged formulation), 
and no such analytical model has appeared yet. 
The phase-dependent calculations of \citet{fbs+09} did not 
include a specific geometrical model for the location of 
the accreting magnetic pole relative to the spin axis, nor any 
treatment of general relativistic effects in the strong 
gravitational field. 

Subsequent papers by \citet{fcrt12} and \citet{ffbb16} 
presented a more detailed approach. 
\citet{fcrt12} adopted the overall BW formulation of the transport
equation but added an additional term to the stochastic part 
of the transport equation intended to account for the effects 
of the bulk motion of the electrons, and these authors also 
implemented a more general power law velocity profile, 
in addition to the simple velocity law given by Equation (2).
However, the seed photons in the \citeauthor{fcrt12} case were 
blackbody distributed across the column with an exponential 
dependence on height.  
The additional model terms are not a part of the original 
BW model, and as such, the \citeauthor{fcrt12} study is not 
an exact implementation of the BW model. 
The inclusion of the additional terms renders the transport 
equation unsolvable using analytical methods, and 
therefore the authors obtain a series of numerical solutions. 
\citet{ffbb16} further enhanced this treatment by including 
a distributed bremsstrahlung and cyclotron seed photon source 
for the Comptonization process. 
Furthermore, \citet{ffbb16} introduced the vertical dependency
of the magnetic field emission in their cyclotron seed 
photon source term.
Moreover, both \citet{fcrt12} and \citet{ffbb16}
fix the average scattering cross sections to
$\sigma_{\parallel} = 10^{-3}\sigma_{\rm T}$ and 
$\bar{\sigma} = 10^{-1}\sigma_{\rm T}$, and therefore the remaining model 
parameters, such as the polar cap radius ($r_0$) and the 
Comptonizing temperature ($kT_e$), are not easy to compare
to our fitted values.
The \citet{ffbb16} treatment was utilized to fit the phase-averaged 
spectral data for the three sources \herxone, 
\fouruooneonefive, and \cenxthree.

Each of the previous models discussed above is an 
important contribution in its own right, and they 
are all offshoots of the original BW model. 
However, each of these models is unable to account for 
three important issues that may have a significant effect 
on the emergent spectra in accretion-powered X-ray pulsars. 
The first issue is the importance of utilizing a self-consistent 
velocity profile in the accretion column. 
This is a central concern, because the velocity field in 
luminous X-ray pulsars is mediated mainly by the radiation pressure, 
which decelerates the flow to rest at the stellar surface. 
But because the radiation pressure profile depends on the 
velocity field through the transport equation, 
ideally, the velocity field calculation should be 
accomplished in a self-consistent manner. 
The self-consistent calculation requires the utilization of a 
sophisticated, iterative algorithm that is beyond the realm of 
implementation in XSPEC. 
The second issue is that none of the cited models utilizes a 
realistic energy equation to determine the vertical variation 
of the electron temperature in the column, instead assuming 
(as we do) that the electrons are isothermal. 
Again, a self-consistent calculation of the thermal structure of 
the column is difficult to include in an XSPEC module. 
Finally, the third issue is the implementation of a cylindrical 
geometry for the accretion column, and a constant magnetic field, 
which is probably acceptable for ``pill-box'' situations, but 
may lead to unknown errors in sources where the column has a 
significant vertical extent, compared with the radius of the star. 
The lack of self-consistency in the treatment of the hydrodynamics, 
the electron temperature, and the accretion geometry clearly 
introduces errors that are very difficult to estimate with precision.

\subsection{Model Limitations and Uncertainties}
\label{subsec:modlimuncer}

The BW model assumes that all of the emergent radiation escapes through
the sides of the magnetic funnel in a fan beam, rather 
than out of the top of the shocked plasma region in a pencil beam. 
Theoretical models available to-date do not settle the question of 
the predominance of either fan beam emission or pencil beam emission 
for accreting X-ray pulsars generally. 
However, some guidance can be obtained from observations.
For example, \citet{leah04} performed a fit to fan and pencil beam 
components of the \herxone\ light curve in the 9$-$14 keV energy range 
during the main high state of the 35-day super-orbital period.
For an assumed distance of 5 kpc, \citeauthor{leah04} found 
that the best fit to the data yielded fan-beam 
and pencil-beam components 
with luminosities of $9.7 \times 10^{35}$ ergs s$^{-1}$ 
and $1.4 \times 10^{35}$ ergs s$^{-1}$, respectively, for 
one assumed pole. 
This is roughly a 7:1 ratio of fan emission vs. pencil emission.
The energy range treated by Leahy is narrower than that we have 
considered above, and thus the \citeauthor{leah04} luminosity will 
not match our observed luminosity. 
However, it does suggest that for \herxone\ at least, the 
assumption of fan-beam-only emission will not result in a large error.

The errors we give in Table 1 for the model input parameters 
result from the formal procedure of asking XSPEC to vary the
likelihood statistic for each parameter to derive the 90\% statistical
error for each parameter in turn.
However, in implementing the BW analytical 
model in XSPEC we have incorporated their assumption that a
simple one-electron-temperature post-shock plasma emission 
region is an adequate approximation of a real flow. 
This results in our using the crude linear velocity relation to 
ensure that the flow stagnates at the NS surface. 
A real model must self-consistently solve for both the dynamical 
structure and the radiation transport simultaneously. 
Such models are in development but are not at the stage where 
they can be implemented in XSPEC (see West et al. 2016, submitted). 
The development of numerical models will also help to answer the
questions regarding the height-dependent seed photon 
injection rates we now discuss.

The height of the sonic point $z_{\rm sp}$ above the 
NS can be estimated from BW Equation (31) 
while the value of $z_{\rm max}$ from BW Equation (80) 
determines the extent of the spatial integration 
of the injected seed photon distribution within the 
accretion column in our model (see Equation (126) of BW).
For the fitted model shown in Table 1, we find 
$z_{\rm sp} = 3.63$ km and $z_{\rm max} = 6.67$ km, 
and hence the sonic point is located within the 
seed-photon integration region as it should be.
The corresponding model optical depth for $z_{\rm max}$ 
we find from BW Equation (79) is 1.33, which compares very favorably 
with the value of 1.37 found originally by BW (see Table 2 of BW).

When we calculate the magnetic field ``blooming" for the
dipole geometry from the NS surface to $z_{\rm sp}$ we find 
that $((R_{\rm NS}+z_{\rm sp})/R_{\rm NS})^3 \sim 2.5$ suggesting 
that our assumption of cylindrical geometry is not very accurate.
Furthermore, the derived magnetic field strength is reduced 
from $B_{12} = 4.25$ at the NS surface to $B_{12} = 1.68$ 
at the height $z_{\rm sp}$ above the NS surface. 
However, this effect is somewhat mitigated by the
fact that cyclotron excitation (and therefore the 
resulting spontaneous emission) is 
concentrated in the dense lower region of the column below $z_{\rm sp}$.
The assumption of cylindrical geometry is not likely to strongly
distort the results relative to a dipole geometry since
most of the radiation escapes through the column walls in
the lower region of the column, in the vicinity of the sonic point 
for the standing, radiation-dominated shock wave \citep{beck98}.

We can understand this by noting that the number of seed photons injected via cyclotron 
decays varies as ${\dot{n}}^{\rm cyc}_{\epsilon} \propto \rho^2 B_{12}^{-3/2}$ \citep{akl87}.
This reflects the fact that excitation to the first Landau level
for the electrons is a two-body process, and also that as the magnetic field 
strength goes up it becomes more difficult for collisions to 
supply the energy needed to excite electrons from the ground state 
to the first Landau level.
At the top of the accretion column ($z = z_{\rm max}$), the local magnetic 
field strength is lower than the nominal value associated with the CRSF, 
which reduces the energy spacing between the ground state and the first Landau 
level making electron excitation easier to accomplish.
However, this is offset by the relatively low value of the plasma density
at $z_{\rm max}$, where the flow velocity is roughly half the speed of light.
These two effects combine to produce a low value of the cyclotron seed photon
production rate ${\dot{n}}^{\rm cyc}_{\epsilon}$ at the top of the column.
Further down in the column, the magnetic field strength is higher, and 
this makes collisional excitation more difficult.
However, this difficulty is overwhelmed by the
rapidly increasing density as the flow is decelerated.
Utilizing our rough velocity parameterization and assuming 
dipolar field geometry we find that for our \herxone\ model, ${\dot{n}}^{\rm cyc}_{\epsilon}$
is strongly peaked at altitudes below $z_{\rm sp}$.
As we noted above we incorporate the BW assumption that the 
cyclotron seed photon production rate is determined by the value 
of the magnetic field strength that is input to the model based 
on the fitted energy centroid of the gaussian CRSF, and the 
effects of Landau level collisional de-excitation are not included.
Future models will need to address how to optimally account 
for the variation in the magnetic dipolar accretion geometry 
and the varying density throughout the column.

Another source of possible error is that we do not include the 
effects of general relativity (GR) in our model.  
GR would redshift the observed luminosity of the accretion flow 
compared to its value at the NS surface and thus increase the 
derived mass accretion rate (at the NS surface) for a given observed 
luminosity. 
For our standard NS parameters the red shift at the NS surface is 
roughly $z_* \sim 0.3$. 
However, for emission further up the column the apparent
red shift would be reduced for emission originating well
above the NS surface.
Thus, it is not as simple as just multiplying 
luminosities and mass accretion rates by factors of $(1+z_*)^2$ and
$(1+z_*)^{-2}$, respectively.
As a pole spins into and out of view that portion of its emission that
is bent around the NS and reaches the observer will vary with phase.
We reiterate that the value for the mass accretion rate we 
derive here is based on full energy conservation (i.e.,  
assuming 100\% efficiency in converting flow gravitational potential
into radiated energy), which is necessary in order to maintain 
consistency with the surface boundary condition (zero energy flux) 
assumed by the BW model.

Another source of error in the application of these calculations
stems from the ambiguity in determining if one pole or
two poles are actually accreting simultaneously in the modeled system.
If only one pole on the star is accreting, then the parameters 
we derive in our fit to the NuSTAR spectrum should be a relatively 
accurate reflection of the real conditions at that pole. 
If, on the other hand, there are two accreting poles, then what we 
observe from Her X-1 is only the phase-averaged luminosity and we 
cannot discern how that luminosity is apportioned between these 
two poles (assuming a dipolar field) using our model. 
Furthermore, the apparent flux coming from each pole is a function 
of spin phase. All we can really say is that if the accretion 
flow onto the NS is substantially divided between two poles, 
then we expect that emission from each pole will result from a 
shock structure that reflects a lower accretion rate than the 
one we adopt here. 
In summary, we can say that the mean accretion rate onto the 
NS is no lower than our fitted rate but we cannot say how 
that accreting material is divided between the accretion poles.

This suggests it would be useful to determine if the character 
of the solutions might change if we relax our energy conservation 
demand and perturb the accretion rate.
Thus, we fit models to the NuSTAR data in which we arbitrarily 
change the accretion rate by $\pm$30\% 
(to $\dot{M} = 1.82 \times 10^{17} \, {\rm g} \, {\rm s}^{-1}$ and 
$\dot{M} = 3.37 \times 10^{17} \, {\rm g} \, {\rm s}^{-1}$, respectively) 
and only demand that the flux in the 5$-$75 keV band is correctly 
accounted for during the fits.
The fit statistics are similar to our base model (see Table~\ref{tbl:herx1fitparam}) 
in that $\chi^2/{\rm DoF} = 549.41/450$\ (-30\%) 
and  $\chi^2/{\rm DoF} = 544.77/450$\ (+30\%) so the 
quality of these fits are all similiar.
We find that the polar cap radius and the scattering cross 
sections do noticeably change, but other fitted parameters, such 
as the Comptonizing temperature, do not change significantly.
For example, the accretion cap radius changed 
from $\sim$107 m with our base model accretion rate
($\dot{M} = 2.59 \times 10^{17} \, {\rm g} \, {\rm s}^{-1}$, 
see Table~\ref{tbl:herx1fitparam}) to $\sim$57 m 
at the reduced rate and $\sim$167 m at the increased rate.
This illustrates that the radius of the accreting pulsar cap is a
strong function of the accretion rate in the BW model.
Moreover, the mean scattering cross section changed 
from $\sim3.5\times 10^{-4} \sigma_{\mathrm{T}}$ for our base model 
to $\sim1.9\times 10^{-4} \sigma_{\mathrm{T}}$ at the reduced rate 
and $\sim5.3\times 10^{-4} \sigma_{\mathrm{T}}$ at the increased rate.
However, the Comptonizing temperature ($kT_e$) changed by less than
3\% in either direction with the variation of the accretion rate. 
This means that the accretion rate changes maintained the basic shock structure 
as radiation-dominated with $(\xi,\delta) \sim (1.30,2.33)$ and
$(\xi,\delta) \sim (1.38,1.91)$ for the low and high accretion rates, 
respectively, whereas $(\xi,\delta) \sim (1.36,2.38)$ for our base accretion rate.
Thus, the models with the perturbed accretion rates are still dominated by 
thermal Comptonization over bulk Comptonization by roughly a factor of 2.

\section{Conclusions}
\label{sec:conclusions}

We describe improved spectral modeling of the accreting X-ray pulsar 
Hercules X-1 with a radiation-dominated radiative shock model, 
based on the analytic work of BW.
We perform a detailed quantitative spectral fit in XSPEC using
a physics-based BW model intended for general release.
This results in 
estimates for the accretion rate, the Comptonizing temperature 
of the post-shock radiating plasma in the accretion column 
as it hits the NS surface, the radius of the accretion column, 
the mean scattering cross section parallel to the magnetic field,
and the flow overall angle-averaged scattering cross section
that regulates the thermal Comptonization process.
We obtain a good fit to the spin-phase averaged 4 to 78 keV X-ray 
spectrum observed by NuSTAR during a main-on of the \herxone\ 
35-day accretion disk precession period over the entire energy 
range of the observation with our 
radiation-dominated radiative accretion shock model.
This demonstrates the utility of the BW radiation-dominated 
radiative shock model implementation in constraining the real physical 
parameters of high luminosity accreting X-ray pulsar sources.

\begin{acknowledgements}

The authors thank Dr. Carlo Ferrigno and Dr. Kenneth Wolfram for 
valuable help in formulating
the numerical implementation of many of the analytical expressions.
We thank Dr. Richard Rothschild for a number of stimulating discussions.
We also thank an anonymous referee for a number of insightful
comments that helped improve the manuscript.
This research was supported by the National Aeronautics and Space 
Administration Astrophysical Data Analysis Program
under grant 12-ADAP12-0118.
MTW and KSW are also supported by the Chief of Naval Research.
JW and F-WS are grateful for support by the Deutsche Forschungsgemeinschaft.
This work made use of data from the NuSTAR mission, a project led by the 
California Institute of Technology, managed by the Jet Propulsion Laboratory, 
and funded by the National Aeronautics and Space Administration. 
This research has also made use of the NuSTAR Data Analysis Software (NuSTARDAS) 
jointly developed by the ASI Science Data Center (ASDC, Italy) and the 
California Institute of Technology (USA).

\end{acknowledgements}

%
%
\begin{deluxetable}{ccc}
\tablecolumns{3}
\tablewidth{0pc}
\tablecaption{\herxone\ XSPEC Model Parameters \label{tbl:herx1fitparam}}
\tablehead{\colhead{Parameter} & \colhead{Value} }
\startdata
$N_{\mathrm{H}}$ (cm$^{-2}$) &$1.7 \times 10^{20}$ (Fixed)\\[0.05in]
Distance (kpc) &6.6 (Fixed)\\[0.05in]
$\dot{M}$ (g s$^{-1}$) &$2.5935 \times 10^{17}$ (Fixed)\\[0.05in]
$kT_e$ (keV) &$4.58_{-0.07}^{+0.07}$\\[0.05in]
$r_0$ (m) &$107.0_{-1.8}^{+1.7}$\\[0.05in]
$B$ (Gauss) &$4.25 \times 10^{12}$ (Tied to $E_{\rm CRSF}$)\\[0.05in]
$\sigma_{\perp}$ ($\sigma_{\mathrm{T}}$) & 1.0 (Fixed)\\[0.05in]
$\sigma_{\parallel}$ ($\sigma_{\mathrm{T}}$) &$5.20_{-0.07}^{+0.08} \times 10^{-5}$\\[0.05in]
$\bar{\sigma}$ ($\sigma_{\mathrm{T}}$) &$3.5_{-0.2}^{+0.2} \times 10^{-4}$\\[0.05in]
$E_{\mathrm{CRSF}}$ (keV) &$37.7_{-0.2}^{+0.2}$ \\[0.05in]
$\sigma_{\mathrm{CRSF}}$ (keV) &$7.1_{-0.2}^{+0.2}$ \\[0.05in]
$\tau_{\mathrm{CRSF}}$ &$0.98_{-0.06}^{+0.06}$\\[0.05in]
$E({\rm FeK}\alpha_{\rm n})$ (keV) &$6.61_{-0.02}^{+0.02}$ \\[0.05in]
$\sigma({\rm FeK}\alpha_{\rm n})$ (keV) &$0.26_{-0.04}^{+0.03}$ \\[0.05in]
$\rm A({\rm FeK}\alpha_{\rm n})$\tablenotemark{a} &$0.0029_{-0.0006}^{+0.0006}$ \\[0.05in]
$E({\rm FeK}\alpha_{\rm b})$ (keV) &$6.53_{-0.08}^{+0.07}$ \\[0.05in]
$\sigma({\rm FeK}\alpha_{\rm b})$ (keV) &$0.90_{-0.14}^{+0.21}$ \\[0.05in]
$\rm A({\rm FeK}\alpha_{\rm b})$\tablenotemark{a} &$0.0043_{-0.0005}^{+0.0005}$ \\[0.05in]
C$_{\rm FPM}$ &$1.037_{-0.002}^{+0.002}$ \\[0.05in]
$\chi^2 / {\rm DoF}$ &$545.55/450$ \\[0.05in]
$\rm Flux(5 - 60 \, {\rm keV})$ (ergs ${\rm cm}^{-2}$ ${\rm s}^{-1}$) &$6.96_{-0.01}^{+0.01} \times 10^{-9}$\\[0.05in]
\enddata
\tablenotetext{a}{Normalization units: Photons cm$^{-2}$ s$^{-1}$ in the line.}
\end{deluxetable}

\end{document}